\pdfoutput=1
\documentclass[10pt]{sigplanconf}

\usepackage{hyperref}
\usepackage{amsmath}

\usepackage{pifont}

\usepackage{xcolor}

\usepackage{flushend}
\begin{document}
\toappear{}

\setlength{\pdfpageheight}{\paperheight}
\setlength{\pdfpagewidth}{\paperwidth}



\titlebanner{DRAFT -- Not for distribution}        
\preprintfooter{Submitted to Onward! '16}          

\title{How Are Programs Found?\\Speculating About Language Ergonomics With Curry-Howard}

\authorinfo{Johannes Emerich}
           {}
           {johannes@emerich.de}

\maketitle

\begin{abstract}
Functional languages with strong static type systems have beneficial properties
to help ensure program correctness and reliability. Surprisingly, their
practical significance in applications is low relative to other languages
lacking in those dimensions. In this paper, the \emph{programs-as-proofs}
analogy is taken seriously to gain speculative insights by analysis of
creation habits in the proof-centric discipline of mathematics. Viewed in light
of this analogy, a sampling of mathematicians' attitudes towards formal proof
suggests that the crucial role of intuition and experimentation in programming
tasks may be under appreciated, hinting at a possible explanation of the
challenges rigorously disciplined languages face in practical applications.
\end{abstract}

\category{D.3}{Programming Languages}{}
\category{K.2}{History of Computing}{}[Software]

\keywords
language ergonomics, type systems, programming systems, programs as proofs, propositions as types


\section{Introduction}

The thoughts in this paper grew over the course of a year, out of the opening
remarks of my thesis defense in logic, as well as a series of informal
discussions I have had since starting an \emph{industry job} as a software
developer. While I am in no way sure that what I am saying is authoritative or
true, it has been my impression that my line of thought has been
thought-provoking for those on the receiving end, and it is in this spirit that
I decided to elaborate my thoughts into a form that might potentially reach a
wider audience.

Being a student of logic with a professional and practical interest in computer
programming, I was looking to write my thesis in an area that would gainfully
combine both topics \cite{emerich15}. The topic of type systems for programming
languages was an obvious fit, but I soon faced with a puzzle: As a student of
logic, everything about sound static type systems seemed obviously right,
beautiful and true, and the research papers and languages I worked through made
perfect sense. On the other hand, from my professional work in software
development, I knew how little attention industry paid to the rightness, beauty
and truth of this approach to programming. My puzzlement is certainly shared in
large parts of the programming language theory community: If ``types are the
leaven of computer programming; they make it digestible'' (Milner), they are
light-weight formal methods that automatically prove ``the absence of certain
program behaviors'' (Pierce) \citep[both][]{pierce02}, and we all want to write
correct programs, why would anyone \textit{not} use them?  And especially, why
would big, for-profit companies opt to not use them, or use such perversions of
type systems as to render any guarantees worthless?

Industry practitioners seldom explain their motives but simply vote with their
feet, or if they do explain, they explain in terms incomprehensible to PL
researchers. It is not surprising then that the disregard for advanced type
systems is often explained away by stupidity, carelessness, or lack of proper
education.  Without even wanting to dispute these labels, I started to
wonder whom programming languages are created for. If a technological artifact
is inadequate for its intended target audience and therefore its intended use
case, it is unacceptable to blame the recipients of the artifact.

To illustrate, let us look at another discipline where technological artifacts
are designed for a large audience, and which is often invoked in certain parts
of the software engineering community: architecture.

In the wake of World War II, war-time destruction and the rapid growth of
metropolitan areas necessitated the rapid provision of affordable housing. In
countries all around the world, a solution was seen in modernist \emph{living
machines}.  High-rise buildings for a new kind of communal living were built
with great enthusiasm, and abandoned with great disappointment as many of the
projects became attractors of poverty and living conditions rapidly
deteriorated. New concepts for urban development with dense housing have
incorporated feedback from this failing to create less fragile living
environments for human beings. Another example of well-intended but
ultimately failed architecture is Haus Wittgenstein in Vienna, a modernist
building designed in part by philosopher Ludwig Wittgenstein. Few of his family
members succeeded to live in this house for long, even while remarking on its
beauty and elegance.

Would we find it convincing if the architects of modernist high-rise projects
or Haus Wittgenstein blamed their dwellers for their inability to dwell in
them?  If they declared them too uneducated to live properly? I think not. Why then
should we let programming language researchers off the hook, if their artifacts
fail to meet the needs of the programming community?

What is to be done to learn about the needs of humans when constructing
programs? Ultimately, this is an empirical question, requiring research
and approximation by trial and error. To guide both processes, it is prudent to
start with some judicious guessing. This speculative
text\label{fix:speculative} is an attempt to judiciously guess plausible
hypotheses from an analysis of 20\textsuperscript{th} century mathematics.

Section \ref{programming-image-maths} introduces a formal connection between
mathematics and computer programming, and questions the implied role of formal
proof in mathematical practice. Section \ref{formalism-history} stops at
several points of historical significance in the past hundred or so years in
mathematics, summarizing discussions about the interplay of intuition,
formalism and rigor. Section \ref{the-analogy} looks back at recurring themes
found in said discussions, takes the programs-as-proofs analogy seriously, and
defines some criteria to evaluate programming systems by. In section
\ref{sec:tool-survey} these criteria are applied to a small number of recent
proposals for programming languages and systems. Section \ref{sec:conclusion}
concludes.

\section{Programming in the Image of Mathematics}
\label{programming-image-maths}

Modern day research in type systems for programming languages occurs in
fruitful collaboration with research in type theory and mathematical logic.
Languages like Haskell and OCaml \hypertarget{fix:haskell-ml}{are part of the
lineage of ML}, a language grown out of a system for theorem proving. GHC, the
\emph{Glasgow Haskell Compiler}, supports a variety of extensions to Haskell's
type system, that bring it closer to the powerful dependent type systems found
in proof assistants like Coq and Agda. The connection between programming
languages and mathematical logic is described succinctly in Per Martin-L\"of's
influential \emph{Constructive Mathematics and Computer Programming}
\cite{martin-loef85}:

\begin{quote}
Parallel to the development from low to high level programming languages, there
has been a change in one's understanding of the programming activity itself. It
used to be looked (down) upon as the rather messy job of instructing this or
that physically existing machine, by cunning tricks, to perform computational
tasks widely surpassing our own physical powers, something that might appeal to
people with a liking for crossword puzzles or chess problems. But it has grown
into the discipline of designing programs for various computational tasks,
programs that have to be written in a formally precise notation so as to admit
of automatic execution. [\ldots] It has made programming an activity akin in
rigour and beauty to that of proving mathematical theorems. (This analogy is
actually exact in a sense which will become clear below.)
\end{quote}

The way in which the analogy is ``actually exact'' is by means of a
Curry-Howard-style correspondence between a fictitious programming language
(borrowing from ALGOL, PASCAL and LISP) and Martin-L\"of's theory of types.
What this correspondence really shows exactly is that in an adequate language,
the activity of writing programs is equivalent to the activity of writing a
formal proof\footnote{For a modern introduction and brief history, see
\cite{wadler}.}.

It is of course the unfamiliarity of the computer scientist or logician with
actual mathematical practice that allows them to conclude that this is, without
qualification, a good thing.  If, however, we look at the history of
mathematics in the last hundred or so years, we see that even in mathematics
the role of \textit{formal} proof is contested or at least much more
complicated than commonly assumed.

The monograph \emph{The Mathematical Experience} stems from 1981, close to
Martin-L\"of's 1985 paper. Therein, Hersh and Davis describe ``the most
mathematician-like mathematician''\footnote{As was customary at the time,
this most mathematician-like mathematician is styled as ``he''.} to display
``the discrepancy between the actual work and activity'' and even ``his own
perception of his work and activity'' \cite[p. 34ff.]{hersh81}. I will quote
selectively, though at some length, to illustrate:

\begin{quote}
He rests his faith on rigorous proof; he believes that the difference between
a correct proof and an incorrect one is an unmistakable and decisive
difference. [\ldots] Yet he is able to give no coherent explanation of what is
meant by rigor, or what is required to make a proof rigorous. In his own work,
the line between complete and incomplete proof is always somewhat fuzzy, and
often controversial.

[\ldots]

To his fellow experts, he communicates [his] results in a casual shorthand. ``If
you apply a tangential mollifier to the left quasi-martingale, you can get an
estimate better than quadratic, so the convergence in the Bergstein theorem
turns out to be of the same order as the degree of approximation in the
Steinberg theorem.'' This breezy style is not to be found in his published
writings. There he piles up formalism on top of formalism. Three pages of
definitions are followed by seven lemmas, and, finally, a theorem whose
hypotheses take half a page to state, while its proof reduces essentially to
\hypertarget{fix:definitions-a-h}{``Apply Lemmas 1--7 to definitions \mbox{A--H}.''}
\end{quote}

When asked for a definition of proof, he gives one as ``cleared up by the
logician Tarski [\ldots] and some others, maybe Russell or Peano'', but also
says that ``of course no one ever really \textit{does} it. It would take
forever!'' The less one knows about formal languages and formal logic, the
better, as ``[that] stuff is all abstract nonsense
anyway''.\label{fix:abstract-nonsense}

Now, would our ideal mathematician be happy to do his work in Martin-L\"of's
formal theory of types? One should think not. Then why are we surprised if the
average programmer does not seem to take to an equivalent programming system?

We will, in the following, undertake a brief historical tour of mathematics to
find some clues as to why mathematical practice does not conform to the ideal
of formal proof.

\section{Intuition and Formal Proof in Mathematics}
\label{formalism-history}

We begin our analysis, abruptly, early in the 20\textsuperscript{th} century, with the
foundational crisis in mathematics in full swing. In response to
misunderstandings and embarrassing failures in the face of an ever more
significant mathematics, mathematicians, logicians and philosophers tried to
establish foundations that would confirm and guarantee their understanding of
mathematics as the house of certainty. To restore the endangered reputation of
mathematics as a prototype of the most rigid science, the goal had to be not
only to axiomatize all of mathematics, but also to show that in the chosen
axiomatization contradictions are in general impossible
\citep[p.~411]{hilbert17}. Even Russell's generous endeavor of axiomatizing
all of logic was not sufficient according to Hilbert, as it was still necessary
to establish that every mathematical problem has a solution (no ignorabimus),
that every solution can be verified, and to find a measure for the simplicity
of a proof. To put this plan into action, mathematical concepts and practices
such as propositions and proofs had to be reified into mathematical objects
that could be studied with mathematical methods. The famous rigor of
mathematics was to be applied to mathematics
itself.\label{fix:birth-of-mathematical-logic} The successes and failures of
Hilbert's program are famous, but its reception was mixed from the onset.

\subsection{Poincar\'e on Intuition and Logic}

Of interest for our analysis is the stance of Henri Poincar\'e, who was at once
appreciative of the quest for certainty and skeptical of the emphasis on
formalism. In a 1900 essay, Poincar\'e identifies ``two entirely different
kinds of minds'' among great mathematicians, ``one sort are above all
preoccupied with logic'', advancing ``step by step, after the manner of a
Vauban who pushes on his trenches against the place besieged, leaving nothing
to chance'', and the other are ``guided by intuition and at the first stroke
make quick but sometimes precarious conquests'' \cite[p.~1012]{poincare05}.
Poincare believed it was the ``very nature of their mind which makes them
logicians or intuitionalists'', and while ``[the] first are incapable of
`seeing in space', the others are quickly tired of long calculations and become
perplexed''. But both are ``equally necessary for the progress of science'',
both ``have achieved great things that others could not have done''.

Poincar\'e however was not blind to the progressive shift towards formality and
rigor, even among the ``intuitionalists'', as ``their readers have required of
them greater concessions''. The reason for this shift is the recognition that
intuition ``cannot give us rigour, nor even certainty''. This is why formalism
and logical analysis are necessary to further progress, a kind of cleaning up
and clarifying that prevents illicit jumping to conclusions. But ``intuition
must retain its role as complement, [...] or as antidote to logic'', not just
for the student but also the creative scientist. To see the unity in a
mathematical problem, we need ``a faculty which makes us see the end from afar,
and intuition is this faculty'' (p.~1018). Poincar\'e uses the example of the
concept of a continuous function, which from the image of a continuous mark of
chalk gradually turned into a construction ``irreproachable in the eyes of the
logician''. But even the logician relies on some sort of intuition, only, as in
the case of Hermite, ``the most abstract entities were for him like living
beings'', allowing them to ``perceive at a glance the general plan of a logical
edifice''. ``In rejecting the aid of the imagination, which, as we have seen,
is not always infallible, they can advance without fear of deceiving
themselves. Happy, therefore, are those who can do without this aid! We must
admire them; but how rare they are!''

Poincar\'e, it seems, was not among them. His thesis adviser characterized him
as follows: ``It must be said, if one wants to give an accurate idea of how
Poincar\'e worked, that many points of [his thesis] needed correction or
explication. Poincar\'e was an \textit{intuitif}'' \cite[cited
after][]{mclarty}. Mclarty states that many of Poincar\'e's publications
offered crucial new insights, laying the foundations to whole fields, all while
employing hasty (or no) proofs, and getting substantial details wrong that were
later to be worked out by others. Poincar\'e's question, ``Who would venture to
say whether he preferred that Weierstrass had never written or that there had
never been a Riemann?'', should be restated for the reader: ``Who would wish
that there had never been a Poincar\'e?''

\subsection{New Math: Real Mathematics Comes to Schools}
\label{sec:new-math}

The foundational crisis in the past, and its shocking revelations absorbed or
ignored, parts of the formalists' results and \textit{spirit} found their way
into research mathematics. Probably the prototypical example of 20\textsuperscript{th}-century
formalist mathematics is the highly influential work of Nicolas Bourbaki, a
group of predominantly French mathematicians. According to a member, Bourbaki
was set up almost in opposition to the \textit{intuitif} Poincar\'e, against
the older generation that ``had learned mathematics in the old-fashioned way'',
``it was not the fashion to value Poincar\'e at all''
\cite{bourbaki-interview}.  In the 1960s, this spirit of renewal found its way
into primary education in mathematics. Today still known (and often ridiculed)
under the moniker ``new math'', various educational reform programs in the
United States and Europe put set theory, Boolean algebra and further \emph{abstract
nonsense} on schools' curricula. A cautionary tale of educational reform
implemented in haste and abandoned without much analysis or care, new math
disappeared as quickly as it had been put on the agenda, as teachers felt
unable to teach and parents unable to understand the new materials.  The
history of the new math is complex and frequently misconstrued
\citep[p.~145]{phillips15}, but the reactions of contemporary mathematicians to
its introduction provide interesting insight into their \textit{attitudes}.
Especially in the United States, the reform efforts led to controversies among
professional mathematicians and scientists, triggered not by the new emphasis
on and interest in mathematics education, but by the image of mathematics
thought to be underlying the reform proposals \cite{phillips14}.

Physicist Richard Feynman in a commentary on new math stresses the importance of
freedom, experimentation and intuition in learning and practicing mathematics
\cite{feynman65}.  While this spirit ``does not appear in [a mathematician's]
final proofs, which are simply demonstrations or complete logical arguments
which prove that a certain conclusion is correct'', it is present in the way
that he works in ``in order to obtain a guess as to what it is he is going to
prove before he proves it''. To give authority to his claim, he quotes a pure
mathematician, J.B.  Roberts:

\begin{quote}
The scheme in mathematical thinking is to divine and demonstrate. There are no
set patterns of procedure. We try this and that. We guess. We try to generalize
the result in order to make the proof easier. We try special cases to see if
any insight can be gained in this way. Finally -- who knows how? -- a proof is
obtained.
\end{quote}

Mathematician and educator Morris Kline spent considerable effort arguing
against the new curriculum, publishing both an essay, \textit{Logic Versus
Pedagogy} \cite{kline70}, and a book, \textit{Why Johnny Can't Add}
\cite{kline74}. His protest against the reform was motivated by a strong belief
in the importance of intuition and gradual development in mathematical training
and practice. The arguments in both texts make much use of historical
developments and anecdotes, illustrating the imperfections and intuitive leaps
present in the work of accomplished mathematicians. Kline presents as an
example the development of calculus from the basic concept of ``instantaneous
rate of change of a function'' \citep[p.~267]{kline70} to the modern formally
precise expression. He documents the vague and faulty conceptualization evident
in the writings of Newton and Leibniz, and Newton's defense of his work against
``overprecise critics'' posing a threat to the ``fruits of invention''. The
same aspect is highlighted in Cauchy's use of differentiability where he had
only assumed continuity, making, on the whole, substantial progress.

Like Feynman, Kline sees mathematics as ``primarily a creative activity, and
this calls for imagination, geometric intuition, experimentation, judicious
guessing, trial and error, the use of analogies of the vaguest sort, blundering
and fumbling'' (p.~271). To counter the response that intuition plays this
important role only in those new to a subject matter, Kline recounts the
anecdote of ``the professor who was presenting a logical proof to his class,
got stuck in the course of the proof, went over to the corner of the blackboard
where he drew some pictures, erased the pictures, and then continued the
proof'' (p.~280).

In the end Kline concedes, quoting Weyl, that ``logic is the hygiene which the
mathematician practices to keep his ideas healthy and strong''.

\subsection{The Threat of Speculative Mathematics}
\label{sec:jaffe-quinn}

Our third episode is set in 1993--1994, a discussion held in the \emph{Bulletin
of the American Mathematical Society} and triggered by a call for action by
Arthur Jaffe and Frank Quinn \cite{jaffe93}. The authors worried about a
perceived trend of an increase in speculative mathematics, due in part to a
cultural clash between mathematics and physics. A nearly categorical statement
opens the paper: ``Modern mathematics is nearly characterized by the use of
rigorous proofs''. This is qualified to say that ``information about
mathematical structures is achieved in two stages'', in the first stage
``intuitive insights are developed, conjectures are made, and speculative
outlines of justifications are suggested'', in the second ``the conjectures and
speculations are corrected; they are made reliable by proving them''. The
two-stage process is compared to physics, divided in theoretical and
experimental physics. For Jaffe and Quinn, theoretical physics is analogous to
the speculative, intuitive stage in mathematical discovery, while experimental
physics is analogous to verification and proof of speculation. Where in physics
there has been productive division of labor along those lines, Jaffe and Quinn
decry what they see as an onslaught of speculative mathematics (which they call
\textit{theoretical mathematics}), arguing that the culture of mathematics is
unprepared. A list of cautionary tales of speculative mathematics gone wrong is
followed by a problem characterization: Speculative work easily goes astray for
lack of corrections from rigorous proof, it hinders further work by causing
confusion about which parts are reliable, it gives glory to the theorizers
while leaving ungrateful cleanup work for others, and finally confuses
newcomers. They end with a short list of \emph{prescriptions} to amend the
situation, mostly calling for explicit labeling of speculative work, and thus
shifting culture to reserve some glory for the task of rigorous validation.

This call was answered in a later issue of the bulletin \cite{jaffe-responses},
in good numbers by some of the speculative \emph{perpetrators} Jaffe and Quinn
had reprimanded. I can only recommend reading the responses, as they provide a
fascinating insight into the varieties of approaches and self-images among
researchers in mathematics. The reader following this advice will notice that
my quotations in the following are highly selective, which should not be reason
for concern, considering that my claim is not that every mathematician looks at
intuition and formal proof in the way suggested, but only that at least
\emph{some} amount of accomplished mathematicians do.

We find the first commentator, Michael Atiyah, ``agreeing with much of the
detail'', but rebelling ``against their general tone and attitude'', presenting
``a sanitized view of mathematics which condemns the subject to an arthritic
old age'' (p.~178). ``But if mathematics is to rejuvenate itself and break
exciting new ground it will have to allow for the exploration of new ideas and
techniques which, in their creative phase, are likely to be as dubious as in
some of the great eras of the past. Perhaps we now have high standards of proof
to aim at but, in the early stages of new developments, we must be prepared to
act in more buccaneering style''. Providing an example, ``Hodge's own proof was
essentially faulty because his understanding of the necessary analysis was
inadequate. Correct proofs were subsequently provided by better analysts, but
this did not detract from Hodge's glory''.

In a scathing comment, Armand Borel refers to the periodic waves of innovation
and rigorization/systematization that have been a permanent feature of
mathematics, accompanied constantly by fears that one might overpower the
other. ``Of course, [...] no part of mathematics can flourish in a lasting way
without solid foundations and proofs''.

Benoit Mandelbrot finds ``most of it appalling''. For him, Jaffe and Quinn's
proposal is reminiscent of the shunning of great intuitive mathematicians such
as L\'evy and Poincar\'e by the mathematical establishment. He even goes so far
as to ask why there had been so few great intuitive researchers of recent, and
speaks of a ``flow of young people'' who were ``acknowledged as brilliant and
highly promising; but they could not stomach the Bourbaki credo'' and left
mathematics.

Saunders Mac Lane responds with an anecdote about a discussion between Atiyah
and himself, ``about how mathematical research is done''.
\begin{quote}
For Mac Lane it meant getting and understanding the needed definitions, working
with them to see what could be calculated and what might be true, to finally
come up with new ``structure'' theorems. For Atiyah, it meant thinking hard
about a somewhat vague and uncertain situation, trying to guess what might be
found out, and only then finally reaching definitions and the definitive
theorems and proofs. This story indicates that the ways of doing mathematics
can vary sharply, as in this case between the fields of algebra and geometry,
while at the end there was full agreement on the final goal: theorems with
proofs. Thus differently oriented mathematicians have sharply different ways of
thought, but also common standards as to the result.
\end{quote}
If only the same could be said about differently oriented programmers! (Mac
Lane goes on to say: ``The sequence of the understanding of mathematics may be:
intuition, trial, error, speculation, conjecture, proof. The mixture and the
sequence of these events differ widely in different domains, but there is
general agreement that the end product is rigorous proof---which we know and
can recognize, without the formal advice of the logicians.'')

We see a wide spectrum of opinions on the role of rigorous proof in
professional mathematics. On one end, Mac Lane voices a clear stance
\emph{against} ungrounded speculation and a demand for proof as the golden
standard, but paired with a laissez-faire attitude towards the \emph{creative}
habits of individuals. On the other, Mandelbrot finds it sufficient to put
forward his discoveries as conjectures, arguing against a unified narrow
conception of what is acceptable mathematics. What we find in common among all
cited here, however, is insistence on the value and necessity of a multitude of
intuitive approaches to mathematical creation.

\subsection{Interactive Theorem Proving}

We shall finally look at some contemporary trends in mathematics. A current
issue that has been in the making for much of the twentieth century but only
recently has found increased attention is the use of computer-aided proof
environments to develop and verify completely rigorous formal proofs. With
this, we loop back to Martin-L\"of's theory of types discussed in the very
beginning.

Our source is the recently published \textit{Type Theory and Formal Proof}
\cite{geuvers}. The book builds up a formal system $\lambda D$ somewhat similar
to Martin-L\"of type theory, extending the Calculus of Constructions with
constructs first introduced in the Automath system of N.G.  de Bruijn. After
having built up the system to a substantial degree, the \emph{real} work
begins, by formalizing a real mathematical theorem (B\'ezout's lemma) and its
proof in the system, illustrating that it is adequate for capturing serious
mathematical content. It may be noted that this shows the adequacy for
formalizing an \emph{existing and known} proof.

Chapter 16, \textit{Further perspectives} uses observations from the proof of
B\'ezout's lemma along with general considerations to reflect on the system
$\lambda D$ and type-theory-based proof assistants more generally. We collect
here some quotes and discuss the implied context under which proof assistants
may be used\label{fix:money-quotes}. ``The type theory $\lambda D$ provides a
system in which mathematical definitions, statements and proofs can be
completely spelled out in a very structured way that is still close to ordinary
mathematical practice.  This enables and facilitates the formalization of
mathematics and the checking of its correctness. [\ldots] The high level of
precision of type theory greatly improves the level of correctness of the
formalized mathematics: incomplete proofs, or proofs using illegal logical
steps, are not accepted'' (p.~379). The diagram on page 381 is relevant,
because it uses as \emph{inputs} informal proof $p$ and informal statement $A$,
thus assumes that both have been formed at this point. If however we have not
yet fully formed $p$ and $A$, the fact that the ``precision of $\lambda D$
guides the proof development'' (p.~380) may come back to bite us. It is no
doubt a good thing that at some point a precise guide will point to flaws in
the details of the development, but doing so early may lead the user astray
trying to verify a nonsensical side-show statement that could easily be
falsified by a quick series of trial and error.  The author has made this
experience first-hand working with the Coq proof assistant.

Their chapter closes with a prognosis: ``Proof assistants have not yet
developed into a standard tool for mathematicians, but we strongly believe they
will in the future'' (p.~387). In the world of mathematics, there is a culture
delineating the phases of discovery and rigorization that suggests methods for
the discovery process and \emph{may} indeed enable publications to spell out
results for proof checkers in ways similar to how results are elaborated into
semi-formal proofs today. We have seen that there is by no means agreement on
this, but at least mathematicians don't generally assume they have to start
with the formal process.

In programming however, I am not aware of a general agreement that problems
have first to be solved in pen and paper before their ``formalization'' as a
program is to be started. What is more, programming tasks frequently are
embedded in real-world interactions that require experiments instead of just
pen-and-paper simulations.

\section{Learning by Analogy}
\label{the-analogy}

We can summarize as follows: The 20\textsuperscript{th} century has seen increased interest in the
foundations of mathematics and a better understanding of the notion of proof. While \emph{rigorous}
proof can be seen as emblematic of modern mathematics, completely \emph{formal} proof has been a
polite fiction or idealization in research mathematics. A vocal community of mathematicians insists
on the continued importance of intuition for the further development of mathematical content and
objects to a prescriptive straitjacket that they fear would stand in the way of creative discovery.

Just how does this affect the practice of computer programming? Like mathematical argumentation,
program construction has undergone a series of \emph{rigorizations} to prevent the final product from
\emph{going wrong}. The parallel is so pronounced that there is an exact formal correspondence between
various formal logics and programming languages, and there is an important sense in which program
construction is equivalent to proof construction. More precisely, it is equivalent to the construction
of \emph{completely formal proof}.

We have seen that in the mathematical community the activity of formal derivation is not generally
seen as adequate for the creation of new mathematical content. This begs the question: How do
\emph{programmers} working in rigorous languages \emph{find} programs?

Mathematicians, in my layman's eyes, have highly idiosyncratic methods of
discovery, and there is no reason for programmers to be any more constrained.
Indeed, when exempt from\label{fix:exempt} corporate restrictions, programmers
\emph{are} free to employ whichever methods they find helpful in discovery:
whiteboard sketches, pen and paper traces, contemplation of denotational
semantics, Node.js prototypes, etc.  And no doubt, in the era before
time-sharing and personal computers, offline algorithm design was the norm, and
one can still find industry veterans today who claim to spend weeks developing
programs without compiling.  Today's generation of programmers however grew up
on REPLs and the edit-refresh-run loop of browser development. The expectation
is that the computer is an interactive device providing instantaneous feedback.
The computer acts as a laboratory for the discovery of programs.

It would certainly be easy to dismiss this approach as dilettantism, the mark
of the amateur, and it is certainly true that experienced practitioners develop
higher level reasoning for familiar problems. But it should be recalled that in
order to understand a problem\label{fix:get-a-grip}, mathematicians employ
``experimentation, judicious guessing, trial and error, the use of analogies of
the vaguest sort, blundering and fumbling'' (Kline) and run through special
cases (Roberts) in their blackboard or pen and paper labs. With a powerful
computing machine on their desks, why would programmers \emph{not} want to make
use of it for exploring the problem space by means of experiment?

Quickly observing the behavior of a few specific instances can give a feeling for the active forces in
a process. Being able to push a value through my program and see the resulting crash can be
illuminating with an immediacy not found in a lifeless compile-time error. For a quick sanity check
I can run a partial implementation early, even though I have not announced all my assumptions
(invariants) to the compiler, \emph{because I know which precise input I will use}. If these
activities are useful and we deprive the programmer of such means of discovery, she will opt for a
system in which the balance between discovery and comprehensibility is more to her (present)
advantage.

\subsection{A Bad Proposal: Prototype in Lisp, Implement in Haskell}

One might think that we already have all the ingredients: Permissive dynamic languages and safe
statically typed languages. So it is fine to fumble and blunder in Python or Lisp, but the real work
needs to happen in a disciplined language. This does not seem to lead very far, however.  The
current trend in software development is towards agile and away from waterfall methods. Programs are
not \emph{discovered} in one creative act to then be reconstructed cleanly and rigorously. Software is
often long-lived, undergoing constant evolution. A possibly stable and \emph{rigorous} core needs to be
seamlessly integrated with a more provisional, in-flux surface area.

It is often an additional requirement to be able to interact with existing
software libraries, such that a separate prototyping environment becomes
unwieldy.  When asked about the motivation to replace the \emph{Structure and
Interpretation of Computer Programs} course with a Python-based course in the
MIT undergraduate curriculum, Gerald Sussman responded that the type of
engineering required to write software had changed in the 1990s. Instead of the
\emph{analysis and synthesis} view taught in \emph{SICP}, what was now needed was
a more experimental, science-like approach in which ``you grab this piece of
library, and you poke at it [\ldots] see what it does'' \cite{sussman}.
Sussman readily admitted that the previous curriculum was more coherent, but
engineering had changed, and it was necessary to find a new way to do and
teach it, even if they were still in search of the right approach.  Not just
first-in-class teaching curricula, but also first-in-class programming systems
need to adapt to the real contextual usage requirements.

\subsection{Gradual Rigor}

If rigorous methods in practical software construction are to succeed, they can not be hermetically
sealed off from environments that allow for creative discovery and preliminary approximations. An
ideal programming system would allow for a wide spectrum of program construction, granting freedom
for the act of discovery and means for the hygiene of rigorization as an approach crystallizes.

As in mathematics, our goal should be enough rigor to keep our programs healthy and strong, but
enough flexibility to enable us to fumble and blunder when still figuring things out. And like in
mathematics, we should be able to move fluently, competently, and with confidence between both ends
of the spectrum.

\paragraph{Realism about the work of the practitioner}
\label{ref:realism}

In a 1979 paper DeMillo, Lipton and Perlis \cite{perlis79} challenged the
software verification community to move from a standard of perfection to a
standard of reliability, arguing that even in mathematics ``absolute rigor''
had, in fact, not been attained, and even less so in applied engineering
disciplines. They pointed to the supposed infeasibility of full verification by
social processes such as the judging of proof by an expert audience. Type
systems can be seen as a partial response to this problem, where verification
is done automatically by machines instead of a human community. The promise of
dependent type systems is to far extend the reach of possible correctness
guarantees embedded in proof-carrying code \cite{asperti09}.

\label{ref:social-proc}
This line of research seems to provide very apt responses to the problem of
verification described by DeMillo et al., by replacing the manual \emph{social}
verification step by machine verification, based on a codification of the
accepted intersubjective standard.  But this does not absolve of the question
of how the design of languages with an eye towards easy automatic verification
affects the resulting languages and its users as \emph{individuals} for the
purpose of initial creation. In this context, it will not do to give some
principled theoretical account of possibility, but details of language
ergonomics will have to be considered. A notorious
\hypertarget{fix:harper-pamphlet}{polemic} declares that untyped languages are
\emph{unityped}, and thus a mere special case of statically typed languages
(\citealt{harper11}, elaborated in \citealt{harper12}). This clarification is
highly interesting, but in no way accounts for the actual usage affordances of
\emph{static} and \emph{dynamic} languages, respectively.  Harper makes great
points about the potential power of dynamicity safely integrated into a
statically disciplined language, while at the same time giving nothing but
condescending explanations for the popularity of \emph{dynamic} languages.

In mathematics, we saw a folklore belief (or pretense?) that rigorous formal
proof is at the center of the mathematician's practice, but considerable
evidence that this is not the full story, and even found vocal individuals
insisting on the importance and legitimacy of non-formal methods. Maybe the
work of mathematicians is \emph{in principle} equivalent to formal proof, but
in the reality of the details, it is not. Maybe also, in the reality of the
details, programmers find it easier to discover solutions to their tasks in the
degenerate unityped languages they choose over languages with strong advanced
type systems.

\subsection{The Landscape Today}
\label{sec:landscape-today}

In today's mainstream programming language landscape, programmers have few
options other than making a choice between languages with a simple and
comprehensible mathematical foundation but with a tightly controlled execution
model that disallows or disincentivizes direct experimentation, and languages
which will allow for fumbling and blundering but do not offer a clear path
towards a rigorous formulation with strong static correctness guarantees.

On one end of the spectrum, Haskell and OCaml are popular statically typed
functional languages. Their semantics are relatively simple and amenable to
mathematical reasoning, and they possess powerful static type systems that
are still growing in expressivity. These are the languages that most seem
to correspond to the vision of programming conjured up by Martin-L\"of.

On the other end of the spectrum, scripting languages like Ruby, Python or
JavaScript are conceptually less predictable and do not lend themselves to
mathematical reasoning. The lack of compile-time type checking makes it easy
to experiment and write programs by incremental approximation, but also
prevents compile-time guarantees about program correctness, and disadvantages
the language with respect to tools support.

Languages like Java, C\# or C++ provide static type systems that give them
compile-time checks, superior tooling, and often superior run-time
performance. On the other hand, they are not constructed around principles
that allow for straight-forward application of modes of mathematical reasoning.
They possess type systems that still leave ample room for fumbling and
blundering, but provide little help in eventually leaving that stage.

Especially in the web programming sphere, there have been recent attempts to
retrofit dynamic languages such as PHP, JavaScript and Python with static type
systems, which often leads to idiosyncratic design choices. Dart and TypeScript
sacrifice soundness of the type system to use ``optimistic'' type compatibility
rules, in the interest of making static types easier to use. Runtime errors are
trapped by the underlying dynamic language~\cite{brandt11}. Similarly
\emph{flawed} type systems have appeared before, if with less pride, for example
in Java.\label{fix:unsound}

If we want to take the \emph{programs as proofs} metaphor seriously, not just as
a theoretical construct, but as something that can actually serve as a practical
model for programming, we need to think about ways of combining the best
properties of the various systems, or come up with altogether new ways of
supporting the modern development process.

\subsection{A Speculative Evaluation Framework}

For programming systems that support development all the way from initial
creation to eventual rigorous codification, we expect supporting capabilities
in two phases: discovery and codification.

\subsubsection{Discovery}

The responses to Jaffe and Quinn hinted at the high diversity of cognitive
styles and approaches among mathematicians. While ultimately mathematicians
have a rigorous common standard of communication, there is a lot of freedom for
individuals to work with their preferred methods and amount of guesswork.
The degrees of freedom a technological system can provide may always need to
be augmented with activities external to the system, yet programming languages
could optimize for the level of flexibility possible when already rigorously
formulated parts of a software system interact with experimental parts. We
define three criteria that are likely to be useful in exploring a problem space
and discovering possible solutions by enabling \emph{fumbling and blundering}.

\paragraph{Special cases}

Testing a procedure by applying it to only some special cases is irresponsible,
but trying a basic idea first on regular, then on edge cases is a time-tested
heuristic for developing the first draft of a general procedure. Considering
special cases allows local and concrete reasoning in place of contemplation of
abstract generalities. A programming environment can allow programmers to run
and observe partial implementations on hand-selected inputs, even if the
implementation has obvious holes that have yet to be addressed.

\paragraph{Flow trumps flaw}

An uninterrupted flow of ideas and associations is more important than the
premature attention to flawed details. I need not work out the details of an
approach I discard after five minutes of exploration.

Today's incarnations of rigorous languages are not very close to \emph{absolute
rigor}. In most programs, there are many properties that are not captured by
the type system (list length, effects, co-effects), and I don't have to battle
the type checker on those dimensions.  Nevertheless, there is a variety of
cases where the type checker insists on guarantees for certain properties. The
more the type system conforms to the ideal of preventing programs from \emph{going
wrong}, the more details will have to be explicitly stated and treated.

While the details need to be taken care of as the overall picture clears up, it
is important to develop a high-level understanding and see an idea through
enough to evaluate its overall value, before investing the time to care about
all the details.

\paragraph{Partial execution}

Especially when making changes to an existing codebase, the programmer may want
to test the changes to one part of the program, while ignoring -- consciously or
unconsciously -- that global invariants have been broken. The environment can
allow the execution of partially functional programs, even if there is no global
coherence.

\subsubsection{Codification}

Discovery and experimentation are important, but so is hardening and gradual
refinement towards a shared and practically verifiable standard of correctness.
Not only does this provide better means for creating reliable artifacts, but it
also is the basis for a common language to communicate ideas and judge
implementations.

\paragraph{Crystallization}

As understanding of the programming task and software system crystallizes, the
programmer should have a clear standard of correctness to work towards, and
should constantly be supported to make the assumptions and invariants explicit
and subject to verification.

\section{Tools for Discovery}
\label{sec:tool-survey}

The goals and criteria described in the preceding section set up
desiderata but do not prescribe any particular solution. A basic
question that is left open, is whether we need to think of this as
strictly a programming language problem, or whether it can be framed
more broadly as a programming \emph{system} problem. The idol of
mathematics presents a pattern where the final form of rigorous proof is
fixed (under adequate idealization), but any path that produces this
form is considered legal.

In program construction, the tools for experimentation, discovery and
crystallization can either be built into the language, or be available
in a programming environment that allows to eventually produce a program
in a final \emph{rigorous} form, even if the language itself has no concept
of imprecision.
Both approaches have been explored in prior work, and it makes sense to
discuss examples viewed through the lens of this distinction.

\subsection{As Programming Language Problem}

A programming language that is to support the development process from
experimentation to refinement into a well-understood, rigid form, needs
to possess a powerful type system that can capture the relevant
behavioral properties, but also allows to encode and tolerate a level of
indeterminacy to support the phases where the programmer is unwilling or
yet unable to formally explain their intention in detail.

Without reference to a particular implementation strategy, this general
idea of viewing the static-to-dynamic continuum as a language-internal
problem has already been called for in \cite{meijer04}.

\paragraph{Gradual Typing}

Gradual typing is an attempt to combine the benefits of static and
dynamic type checking, yielding control of which parts of the program
are statically checked to the programmer. At the core, gradual type
systems extend static type systems with a special \texttt{dynamic} type
that is similar to a global supertype (\texttt{Object} in Java) that
also acts as a global subtype. To prevent the type hierarchy from
collapsing under the subtyping relation, however, a new non-transitive
\emph{consistency} relation is introduced that allows any type to be
implicitly converted to and from \texttt{dynamic}.
The \texttt{dynamic} type can be
associated with an expression either explicitly or through omission of
type information, in which case the expression can act as any type at
compile time, with dynamic checks for each operation performed at
run-time.

By supporting both approaches in the same language, gradual typing
enables programmers to evolve programs from an implementation with only
dynamically checked types to a more predictable codebase with
compile-time correctness guarantees. As the program evolves, fragments
without static type checks can enter and be annotated repeatedly, or
even remain part of the code base indefinitely.

The use of \emph{special cases} is facilitated by the generous
type conversions possible in the system, but may require manual effort
on the side of the programmer. The ability to use dynamically typed
fragments is supportive of programmer \emph{flow}. \emph{Partial
execution} is possible, but limited to the dynamically typed fragments
of a program. The presence of a clearly defined and enforced type
discipline provides a framework for \emph{crystallization} but relies
on the programmer's discipline to make use of it.

The term \emph{gradual typing} was introduced in \cite{siek06}.
Recent research has been focused on improving performance and
communication of type errors \cite{garcia16, takikawa15}.
Established languages that support some version of gradual typing are
C\#, or more recently Dart and TypeScript.

\paragraph{Dependent Types}

The question of how much type information to extract from the programmer
becomes more relevant as the expressiveness of type systems grows. In
\emph{Why Dependent Types Matter}, Altenkirch et al. present a series of
refinements of merge-sort written in Epigram, a functional language with
dependent types. In their words: ``It is the programmer's choice to what
degree he wants to exploit the expressiveness of such a powerful type
discipline. While the price for formally certified software may be high,
it is good to know that we can pay it in installments and that we are
free to decide how far we want to go'' \cite{altenkirch05}. Because of
the expressive type system it is possible to encode not just structural
information about function input and output, but more complex
required and guaranteed properties for input and output -- pre- and
post-conditions. As an example, a vector concatenation function may
carry the information that the length of the output vector is the sum of
the input vectors. If the compiler is unable to infer this fact from the
function definition, the programmer may have a proof obligation on her
hands! In this case, as an alternative to providing a proof of this
simple fact, Epigram allows to use a more imprecise type that will
merely express that the output vector has \emph{some} length.

The ability to choose the level of detail of typings grants programmers
the power to move more or less quickly and revisit parts of the code
deserving of greater precision over time. This helps prevent disruptions
of \emph{flow}, and allows for straight-forward \emph{crystallization}.
It is possible to observe execution instances for selected
\emph{special cases} as long as they are permitted under the current level
of type precision. \emph{Execution} of any part of the program is only
possible if the whole program compiles.

\subsection{As Programming System Problem}

\paragraph{Statically-Aided Discovery}

Instead of manually running experiments on code with lacking type
information, programmers can provide type information and receive
automatic and direct feedback from the programming environment that may
help revise and refine types and implementation.

\cite{dybjer} presents an extension to the proof assistant Agda, that
combines static and run-time analysis by means of property-based testing
to quickly identify dead-ends in formal proofs with high probability. In
this manner, the system user can gain confidence in the validity of
her current attempt, before potentially wasting time and energy trying
to prove some ineffable, false property. This approach supports
\emph{crystallization} by reducing the time to feedback. \emph{Flow} is
somewhat disrupted by switching from implementation to verification, but
\emph{examples} for testing are even automatically generated.
\emph{Partial execution} is limited to the fragment under consideration.

A similar system is described in \emph{Dynamic Witnesses for Static Type
Errors} \cite{seidel16}, which aims to complement the abstract error
messages provided by OCaml's type checker with problematic example
inputs and an accompanying trace that illustrate the source of the
error. \emph{Special cases}, that is example inputs and executions, are
provided only for negative cases that produce a type error. Because the
tool is external to the target language and compiler, \emph{execution}
of partially erroneous programs and testing with arbitrary examples is
otherwise not possible. \emph{Flow} is disrupted in just the same way as with a
regular compiler, but the duration of the disruption may be decreased.
There is no added support for gradual \emph{crystallization}, even while
there is a clear and enforced standard of correctness.

There is a tradition of type-directed tools for interactive proof construction
in proof assistants like Agda, Coq and Isabelle. A graphical user interface can
present the available hypotheses and open proof obligations and allow to select
standard routines (\emph{tactics}) to automatically proceed with parts of the
proof construction \cite[see also][]{asperti09, geuvers}. The same idea can be
applied to aided program construction as suggested in \cite{altenkirch05}: The
environment lists the current programming goals based on the available type
information and the programmer either implements them manually or by means of a
meta-programming rule available as library code.

Instead of supporting experimental discovery, the programming environment provides
support in reasoning at the static level, gradually working towards a pre-defined
and checked type-level description of each part of the program.
The language Idris, designed
for general purpose programming, has a compiler with explicit support for this
style of interactive type-directed development~\cite{brady16}.

\paragraph{Pluggable Type Systems}

Type disciplines can be seen as entirely external to the programming
language, in which case there may be not one, but many static type
systems designed for a language, allowing the programmer to pick the
most appropriate one or ones. Such optional, pluggable type systems that
have no effect on language semantics were argued for in \cite{bracha04}.
Because pluggable type systems are essentially analysis tools for
dynamic languages, all discovery-phase benefits for \emph{flow},
\emph{special cases} and \emph{partial execution} transfer unimpinged. On the other
hand, type systems become more of a tool for individual developers and
there is no longer any notion of \emph{crystallization} towards a common
standard of rigor.

\paragraph{Flexible Execution Models for Typed Languages}

This category of approaches is closest to gradual typing in that a
designated type system or standard of rigor is assumed and combined
with the desire to emulate the development experience of dynamic
languages. Unlike in gradual typing, however, the type system is not
enriched to allow the distinction between \emph{static} and \emph{dynamic}
fragments of a program. Instead, the programming environment is altered
to (optionally) defer certain type checks to run-time, allowing the
execution even of provably type-incorrect programs. Execution may fail
with a trapped error if the execution path leads to an error, or
continue without failure as long as no problematic part of the program
is reached.

\emph{Always-Available Static and Dynamic Feedback} \cite{bayne11}
notes that ``[most] statically-typed languages embody the philosophy
that an ill-typed program is of zero value'' and henceforth reject such
programs. Instead, Bayne et al. see such programs as valuable for
experimentation and want to provide feedback from both static analysis
and dynamic execution at any time.

\emph{Progress types} \cite{politz12}, instead of giving the programmer
control over which parts of the program are disciplined by a static type
system, grants control over which type errors are to be compile-time
errors, such that irrelevant errors can be ignored until they occur at
run-time. The authors seem to see this approach as compatible with the
idea of pluggable types, however the presented system is focused on the
existing run-time type system of the base language.

\emph{Liberating the Programmer with Prorogued Programming}
\cite{afshari12} aims at creating a new programming paradigm, based
around a more interactive program creation workflow. Their approach is
based around three principles: allowing the programmer to defer
implementation partially to focus on an ongoing concern; allowing the
programmer to supply appropriate values during execution; and allowing
the programmer to execute partial implementations at any time.

In all three cases, disruptions to \emph{flow} are minimal, \emph{special cases} can
be hand-picked, and \emph{execution} can happen at will. With the possible
exception of progress types, \emph{crystallization} towards a
statically-checked type discipline is an explicit goal and can be pursued
gradually. An early version of the basic idea is described in
\cite{cartwright91}. Recent versions of the Glasgow Haskell Compiler support
the deferral of type checks to run-time \cite{vytiniotis12}.

\section{Conclusion}
\label{sec:conclusion}

There is much work left to be done, both in working out the details of even
these speculative arguments, and if they convince, in empirical investigation
of their implications.  When reading the writings of researchers, one finds
references to the popularity of so-called dynamic languages, but this often
appears as a curious, brute fact, something not available to or worthy of
analysis.  Even publications presenting technologies in support of the
\emph{dynamic habits} of programmers seem to confine themselves to appealing to
only that brute fact, putting the \emph{blame} on programmers instead of
justifying the benefits independently. To give just one example: ``The ethos of
gradual typing takes for granted that programmers choose dynamic languages for
creating software'' \cite{takikawa15}.  My hope is to convince the reader that
there is something to learn from looking at programs as proofs, and \emph{then}
looking at how proofs really come about, without falling prey to idolizing
mathematics.

Looking at the two examples drawn from architecture and urban planning we can
make a distinction that may help address a possible objection. On one side,
Wittgenstein's work raises doubts about whether his design was guided by the
ambition to provide housing\footnote{It brings to mind \emph{his} metaphorical
invocation of architecture: ``I am not interested in erecting a building but in
having the foundations of possible buildings transparently before me''.}. On
the other side, it seems plausible (if not necessary) to believe that the
architects of modernist housing projects \emph{were} in fact motivated by the
desire to provide housing.  Unfortunately, this does not mean that the results
are altogether different in practice.  Similarly, my claim need not be that
language designers are \emph{aiming} for a formalist paradise devoid of
practical motivation. But even if for the best intentions their design takes
inspiration from mathematics in a too simplistic manner, it may miss important
aspects and lead to similarly \emph{formalistic} outcomes.

If we focus too much on making ``programming an activity akin in rigour and
beauty to that of proving mathematical theorems'', while losing track of the
``rather messy job of instructing this or that physically existing machine, by
cunning tricks, to perform computational tasks'', and not even paying attention
to how mathematicians really do prove mathematical theorems, we run the risk of
erecting, like Wittgenstein, a dwelling for the gods. The reality of the
practitioner is not as clean-cut and elegant as its idealization, but makes use
of a variety of mindsets, cognitive modes, and good old fudging. Coming up with
an account of how the rigorous formal systems we have conceived of can
integrate with less rigid components will be challenging, but worthwhile.


\acks{%
  The seed of doubt was planted during my graduate studies at the ILLC and
  further shaped by my work at Prezi. Andras Slemmer's interest prompted me to
  produce an elaborated written version. Jan Martin Raasch, Ignas
  Vy\v{s}niauskas and Christoph Gietl helped by discussing the written version
  at various stages. The reviewers caught some of my sloppier thinking and
  suggested further material and distinctions.
}


\bibliographystyle{abbrvnat}


\end{document}